\documentclass[]{aa}
\usepackage{txfonts,natbib,graphicx,color}
\bibpunct{(}{)}{;}{a}{}{,}

\def\apjl{ApJ}

\def\apjs{ApJS}

\def\aap{A\&A}

\begin{document}

\title{Spin frequency distributions of binary millisecond pulsars}

\author{A.~Papitto\inst{\ref{inst1}}\thanks{papitto@ice.csic.es}~~ \and  D.~F.~Torres\inst{\ref{inst1}}$^{,}$\inst{\ref{inst2}} \and N.~Rea\inst{\ref{inst1}}$^{,}$\inst{\ref{inst3}} \and T.~M.~Tauris\inst{\ref{inst4}}$^{,}$\inst{\ref{inst5}}}

\institute{Institut de Ci\`encies de l'Espai (IEEC-CSIC), Campus UAB,
  Fac. de Ci\`encies, Torre C5, parell, 2a planta, 08193 Barcelona, 
  Spain\label{inst1}
 \and Instituci\'o
  Catalana de Recerca i Estudis Avan\c{c}ats (ICREA), 08010 Barcelona,
  Spain\label{inst2} 
\and Astronomical Institute "Anton Pannekoek", University of Amsterdam, Postbus 94249, NL-1090-GE Amsterdam, The Netherlands\label{inst3} 
\and Argelander-Insitut f\"{u}r Astronomie,
  Universit\"{a}t Bonn, Auf dem H\"{u}gel 71, 53121 Bonn,
  Germany\label{inst4}
 \and Max-Planck-Institut f\"{u}r
  Radioastronomie, Auf dem H\"{u}gel 69, 53121 Bonn,
  Germany \label{inst5}}

\date{Received date }

\abstract{ Rotation-powered millisecond radio pulsars have been spun
  up to their present spin period by a $10^8-10^9{\rm yr}$ long
  X-ray-bright phase of accretion of matter and angular momentum in a
  low-to-intermediate mass binary system.  Recently, the discovery of
  transitional pulsars that alternate cyclically between
  accretion and rotation-powered states on time scales of a few years
  or shorter, has demonstrated this evolutionary scenario. Here, we
  present a thorough statistical analysis of the spin distributions of
  the various classes of millisecond pulsars to assess the evolution
  of their spin period between the different stages. Accreting sources
  that showed oscillations exclusively during thermonuclear type I
  X-ray bursts (nuclear-powered millisecond pulsars) are found to be
  significantly faster than rotation-powered sources, while accreting
  sources that possess a magnetosphere and show coherent pulsations
  (accreting millisecond pulsars) are not. On the other hand, if
  accreting millisecond pulsars and eclipsing rotation-powered
  millisecond pulsars form a common class of transitional pulsars,
  these are shown to have a spin distribution intermediate between the
  faster nuclear-powered millisecond pulsars and the slower
  non-eclipsing rotation-powered millisecond pulsars. We interpret
  these findings in terms of a spin-down due to the decreasing
  mass-accretion rate during the latest stages of the accretion phase,
  and in terms of the different orbital evolutionary channels mapped
  by the various classes of pulsars. We summarize possible
  instrumental selection effects, showing that even if an unbiased
  sample of pulsars is still lacking, their influence on the results
  of the presented analysis is reduced by recent improvements in
  instrumentation and searching techniques.  }


\titlerunning{The spin frequency distribution of binary millisecond pulsars}
\authorrunning{A.~Papitto et al.}

\maketitle

\section{Introduction}
\label{sec:intro}

The accretion of $0.1-0.2$ M$_{\odot}$ transferred by a companion star
through an accretion disc in a low-mass X-ray binary (LMXB) is the
widely accepted process to explain old neutron stars (NSs) that spin
at a period of a few milliseconds
\citep{alpar1982,radhakrishnan1982,bhattacharya1991}. This framework
is known as the {\it recycling} scenario and identifies accreting
X-ray bright NSs as the progenitors of radio-millisecond pulsars whose
emission is powered by their loss of rotational energy.  In this
scenario, after a $0.1-10$~Gyr long X-ray-bright accretion phase that
spins up the NS \citep[see, e.g.,][]{tauris2012b}, the pressure
exerted by the NS magnetic field is able to sweep out the
light-cylinder volume, sparking the rotation-powered pulsed radio
emission of the {\it recycled} pulsar.  Mass-accretion not only spun
up the $\ga 300$ binary and isolated rotation-powered millisecond
pulsars known in the Galaxy, but probably also reduced their magnetic
field strength to values of $10^7-10^9$ G \citep[see, e.g.,][for a
  review]{bhattacharya1995}.

The accretion-driven spin up of an NS in an LMXB is demonstrated by
the observation of coherent pulsations at a period of a few ms from 15
NSs (accreting millisecond pulsars, hereafter accreting MSPs;
\citealt{wijnands1998}, \citealt{patruno2012c}), and of
quasi-coherent-oscillations observed from an additional ten sources
exclusively during thermonuclear type I X-ray bursts (nuclear-powered
millisecond pulsars, hereafter nuclear MSPs\footnote{We stress that
  here, the term {\it nuclear} only refers to the mechanism that
  powers the emission of the source while it shows pulsations, whereas
  for most of the time, the X-ray output is due to mass accretion.};
\citealt{chakrabarty2003}, \citealt{watts2012}). The swings between
rotation and accretion-powered behaviour recently observed from IGR
J18245--2452
\citep{papitto2013nat,pallanca2013,ferrigno2013,linares2014} PSR
J1023+0038 \citep{archibald2009,stappers2013,patruno2014}, and XSS
J12270--4859 \citep{bassa2014,papitto2014,roy2014,bogdanov2014}
demonstrated that NSs in some low-mass X-ray binaries alternate
cyclically between rotation and accretion-powered states on short
time-scales of a few years, or shorter.

In this paper, we compare the spin frequency distributions of the
various samples of MSPs discovered so far, to test theories for describing
the spin evolution of NSs as they evolve from the accretion to the
rotation-powered stages of the recycling scenario \citep[see,
  e.g.,][and references therein]{tauris2012}.

\begin{figure}
 \resizebox{\hsize}{!}{\includegraphics{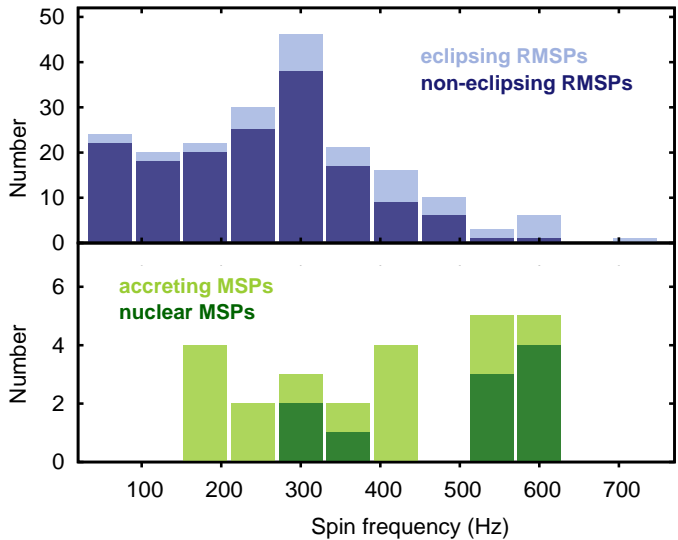}}
\caption{Spin frequency distribution of binary non-eclipsing and
  eclipsing rotation-powered millisecond pulsars (top panel; blue
  and light blue), and nuclear-powered and accreting millisecond pulsars
  (bottom panel; green and light-green).} \label{fig:Freqdistr}
\end{figure}

\section{Population of rotation- and accretion-powered
binary millisecond pulsars
}

To rely on the most recent sample of the rapidly growing population of
rotation-powered millisecond pulsars (hereafter RMSPs), we considered
the lists of radio MSPs compiled by D.~R. Lorimer\footnote{available
  from {http://astro.phys.wvu.edu/GalacticMSPs}} and
P.~C.~C. Freire\footnote{available from
  {http://www.naic.edu/~pfreire/GCpsr.html}}, corresponding to sources
found in the Galactic field and globular clusters, respectively. As of
January 2014, these lists contain 339 radio pulsars with a spin period
shorter than 30 ms. Slightly fewer than half of them are isolated
sources, or systems for which an orbital period could not yet be
determined.  In the following, we only consider binary RMSPs to
compare them with their accreting progenitors, since the evolutionary
history of isolated MSPs might have been somewhat different than that
of pulsars with a companion star
\citep[e.g.][]{verbunt1987,kluzniak1988}.  The binary RMSPs that show
irregular eclipses of their radio emission \citep[][hereafter
  eclipsing RMSPs]{roberts2013} share a close link with their
accreting progenitors. In fact, the observed eclipses are due to
absorption caused by matter lost by the companion star that is ejected
from the system by irradiation of the pulsar
\citep{fruchter1988,kluzniak1988}. Because the semi-degenerate
companion star of at least some of these eclipsing pulsars is very
close to filling their Roche lobe \citep[see, e.g.,][]{breton2013},
the mass lost by these stars may then occasionally overcome the
pressure of the radio pulsar, driving the system into an accretion
state. Switches between accretion- and rotation-powered stages have
recently been observed from three eclipsing RMSPs
\citep{archibald2009,papitto2013nat,patruno2014,bassa2014}. We
summarize the properties of the various populations of binary RMSPs in
Table~\ref{tab:distr} and plot the distribution of their frequencies
in the top panel of Fig.~\ref{fig:Freqdistr}.


\begin{figure}
 \resizebox{\hsize}{!} {\includegraphics{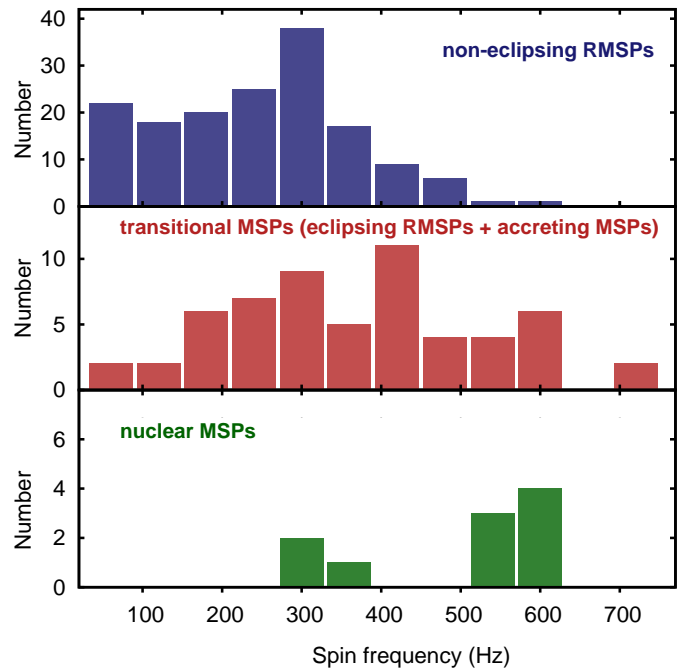}}
\caption{Spin frequency distribution of non-eclipsing RMSPs (blue),
  transitional MSPs (defined as the sum of eclipsing RMSPs and
  accreting MSPs; red) and nuclear MSPs (green).}
 \label{fig:Perdistr}
\end{figure}


\begin{table*}
\begin{minipage}{137mm}
\caption{Samples of binary millisecond pulsars.}
\label{tab:distr}
\centering
\renewcommand{\footnoterule}{} 
\begin{tabular}{@{}l r c c c  l}
\hline\hline
Class  & N & $\langle {\nu}\rangle$ (Hz) & $\tilde{\nu}$ (Hz) & $\sigma_{\nu}$ (Hz) & Description \\
\hline

RMSPs & 199 &  268 & 274 & 139 & Rotation-powered (radio) pulsars in a binary system \\
$\:$ Galactic RMSPs & 128 & 262 & 267 & 141 & RMSPs in the Galactic field \\
\vspace{0.2cm} $\:$ cluster RMSPs & 71 & 277 & 276 & 138 & RMSPs of globular clusters \\
$\:\:$eclipsing RMSPs & 42 & 367 & 377 & 154 &  RMSPs whose radio pulsed emission is irregularly eclipsed  \\
\vspace{0.3cm}$\:\:$non-eclipsing RMSPs & 157 & 242 & 252 & 122 &  RMSPs whose radio pulsed emission is not eclipsed\\

LMXBs & 25 &  413 & 401 & 150 & All accretion-powered pulsars (accreting MSPs + nuclear MSPs) \\
$\:$ accreting MSPs & 15 & 353 & 377 & 140 & Accreting millisecond pulsars (X-ray coherent oscillations) \\
\vspace{0.2cm} $\:$ nuclear MSPs & 10 &  502 & 552 & 123 & Nuclear-powered millisecond pulsars (only X-ray burst oscillations) \\
transitional MSPs  & 56 & 365 & 377 & 150 & Transitional pulsars (accreting MSPs + eclipsing RMSPs)\\ 

\hline
\end{tabular}
 \end{minipage}
\tablefoot{Number of objects, mean, median and standard deviation, of
  the spin frequency distributions of the considered classes of
  rotation and accretion-powered millisecond pulsars ($P<30$ ms) in
  binary systems. IGR~J18245$-$2452 is counted as both an eclipsing- and an accreting MSP.   }
\end{table*}



The spin period is known for 25 accreting NSs in LMXBs (15 accreting
MSPs, 10 nuclear MSPs, cf. Section~1).  All accreting MSPs and seven
out of ten nuclear MSPs are X-ray transients, accreting at a rate up
to $10^{-9}$ M$_{\odot}$ yr$^{-1}$ only during their few-weeks-long
outbursts, and are found in quiescence at levels corresponding to less
than $<10^{-12}$ M$_{\odot}$ yr$^{-1}$ for most of the time
\citep[e.g.][and references therein]{coriat2012}.  Even though these
two classes overlap (burst oscillations were also observed from eight
accreting MSPs), we considered sources that showed oscillations only
during bursts (nuclear MSPs) to form a distinct class, because so far
they did not show evidence of possessing a magnetosphere able to
channel the accreted mass close to the magnetic poles and yield
coherent pulsations. Nuclear MSP bursts tend to be observed in
brighter states that are characterized by a softer spectrum, while
they occur even in the hard state in accreting MSPs
\citep{galloway2008,watts2009,watts2012}, which possibly indicates
that an important role is played by the magnetosphere in determining
burst oscillation properties. The properties of the spin distribution
of LMXB pulsars are summarized in Table~\ref{tab:distr} and are
plotted in the bottom panel of Fig.~\ref{fig:Freqdistr}.

The recent discovery of systems that switch between rotation-
and accretion-powered states on short time-scales also motivated us to
define an intermediate class of systems, dubbed transitional pulsars
(transitional MSPs). These systems behave as eclipsing RMSPs when the
mass in-flow rate is low and the pressure of the pulsar prevents mass
from falling onto the NS, while they are seen as X-ray-bright
accreting sources when the transferred mass overcomes this
pressure. IGR~J18245--2452, observed as an accreting MSP and an
eclipsing RMSP at different times \citep{papitto2013nat}, is the
prototype of this class. A transition to a rotation-powered state
during X-ray quiescence has also been proposed for a number of
accreting MSPs on the basis of the observed spin-down
\citep{hartman2008,hartman2009,patruno2010b,hartman2011,papitto2011,riggio2011,patruno2012b},
reprocessed optical emission
\citep{burderi2003,campana2004,disalvo2008,burderi2009,davanzo2009},
and orbital evolution
(\citealt{disalvo2008,burderi2009,patruno2012b}), even if radio and
$\gamma$-ray pulsations have not been detected so far
\citep{burgay2003,iacolina2009,iacolina2010,xing2013}. On the other
hand, state transitions to an accretion stage have only been observed
from eclipsing RMSPs (see above); these sources are the only possible
candidates to undergo such fast transitions because the companion
star has to spill matter through the inner Lagrangian point of the
orbit to do so. Hence, we define the class of transitional MSPs as the
sum of accreting MSPs and eclipsing RMSPs.  The spin distribution of
this class is plotted in the middle panel of Fig.~\ref{fig:Perdistr},
where the distribution of non-eclipsing millisecond radio pulsars
(non-eclipsing RMSPs) and nuclear MSPs are also shown.

We note that only sources with a spin period shorter than 30~ms are
considered in these samples because we aim at comparing only pulsars
that are at (or close to) the end of the recycling process. While the
exact value of this threshold is somewhat arbitrary, a significantly
shorter period may indicate that the spin-up process has been
interrupted halfway or has recently been activated, which may occur in
the dense environment of globular clusters where dynamical exchange
interactions are relatively frequent \citep{verbunt1987}. This
criterion excludes of an accreting source such as the 11~Hz pulsar
IGR~17480--2446 in Terzan 5, which is thought to be just mildly
recycled, having initiated its spin-up phase shorter than about 10~Myr
ago \citep[see, e.g.][]{patruno2012e}.

\section{Comparison of the spin distributions}
\label{sec:comparison}

To compare the spin frequency distributions of the various classes
defined in the previous section, we relied on the Kuiper test
\citep{kuiper1960,stephens1974,paltani2004}. A Student t-test
\citep[as applied by][]{tauris2012} is inadeguate to compare the
various spin distributions because most of these spin frequency
distributions are very different than a normal distribution, as for
example can be shown with a Shapiro-Wilk test. The Kuiper test is
preferred over the Kolmogorov-Smirnov \citep[used, e.g.,
  by][]{hessels2008}, because it is equally sensitive to deviations
close to the median and towards the tails of the distribution.

The spin distributions of binary rotation-powered millisecond pulsars
in the Galactic field (Galactic RMSPs) and those belonging to a
globular cluster (clusters RMSPs) are similar. The probability that
the two ensembles are drawn from the same distribution is 0.28, and we
therefore considered them as a single sample. On the other hand,
eclipsing RMSPs and non-eclipsing RMSPs have a significantly different
spin distribution. The former group has a mean spin frequency higher
by $\sim\!125$ Hz, and the probability that the two ensembles share
the same parent distribution is $8.5\times10^{-4}$.

 A glance at Table~\ref{tab:distr} and Figs.~\ref{fig:Freqdistr} and
 \ref{fig:Perdistr} might lead one to conclude that accreting MSPs are
 faster than binary millisecond radio pulsars (RMSPs), as noted by
 several authors \citep{ferrario2007,hessels2008,tauris2012}. However,
 even though accreting MSPs are on-average faster than RMSPs by $\sim
 100$ Hz, comparing the two distributions gives a probability of 0.374
 that they come from the same parent distribution, which clearly does
 not allow us to identify any significant difference.  To determine
 the influence of the paucity of known accreting MSPs on this result,
 we simulated $10^7$ samples of $\mathcal{N}$ objects with spin
 frequencies based on a normal distribution with the same mean and
 standard deviation as known accreting MSPs (see
 Table~\ref{tab:distr}). We then varied the number of simulated
   objects ($\mathcal{N}$) and re-run the simulations.   To
   detect a $3\:\sigma$ difference to the RMSP distribution in 90 per
   cent of the $10^7$ simulations we run, there need to be at least
   $\mathcal{N}=50$ accreting MSPs.  We conclude that if the spin
 frequency distributions of accreting MSPs and RMSPs are different, we
 would be able to detect thisdifference only when the number of known
 accreting MSPs has increased by more than a factor of three (assuming
 a normal distribution).

On the other hand, accreting sources that showed pulsations only
during type-I X-ray bursts (nuclear MSPs) are the fastest class of
objects among those defined in this work. Their average frequency is
roughly twice as fast as that of RMSPs and $\sim$90 Hz higher than
that of accreting MSPs. The difference to RMSPs is significant because
the probability that they come from the same distribution is
$2.5\times10^{-3}$.  The probability that the samples of
accreting MSPs and nuclear MSPs come from the same parent distribution
is 0.154, although this value is hardly constraining; we have
performed equivalent simulations as explained above and found that
each sample needs at least 60 objects for a Kuiper test to identify a
statistically significant spin difference between these two classes.

The total LMXB sample of accreting MSPs and nuclear MSPs has a
  higher average frequency than RMSPs, with a probability of only
$\la 1$ per cent that the two samples share the same distribution.

Accreting-MSPs and eclipsing RMSPs have a very similar spin
distribution (see Table~\ref{tab:distr}). The difference between their
average frequencies is just $14$ Hz, and the probability that the two
samples come from the same parent distribution is 0.912. This is
compatible with the hypothesis that they form a common class of
objects at the same evolutionary epoch, which we labelled
transitional MSPs. From comparing transitional MSPs with
non-eclipsing RMSPs we find that the probability that the two
distributions are compatible is $3.9\times10^{-5}$, which indicates a
significant difference. A low probability ($0.048$) is also found when
transitional MSPs are compared with the faster nuclear MSPs. The
tentatively defined class of transitional MSPs therefore has an spin
frequency distribution intermediate between that of nuclear MSPs and
non-eclipsing RMSPs.

\begin{figure}
  \resizebox{\hsize}{!}{\includegraphics{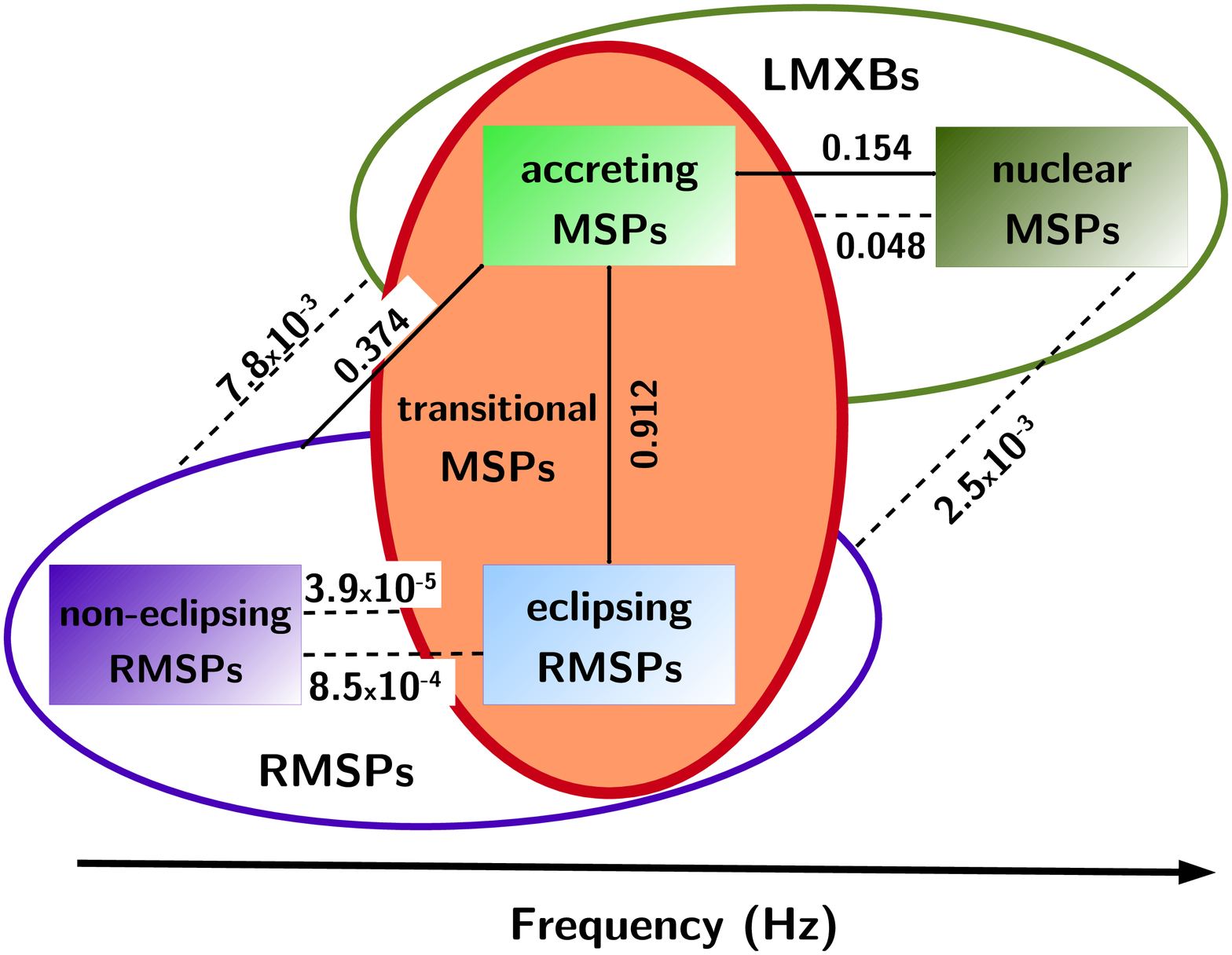}}
 \caption{Relational diagram of the various samples defined in this
   work, based on the comparison of their spin frequency
   distributions. Dashed lines connect samples that have a probability
   lower than 5 per cent of coming from the same parent distribution,
   while solid lines connect those with a higher probability. Numbers
   give the probability that the considered samples are drawn from the
   same distributions, evaluated with a Kuiper test.}\label{fig:rel}
 \end{figure}

We summarize our results in the relational diagram plotted in
Fig.~\ref{fig:rel}, where dashed lines connect samples with a
probability lower than $5\%$ of being drawn from the same parent
distribution, while solid lines connect those for which we evaluated a
higher probability.

\section{Discussion}
\label{sec:disc}

Our analysis shows that nuclear MSPs form the only class of
accretion-powered pulsars that are significantly faster than RMSPs,
while accreting MSPs are not. However, if accreting MSPs and
eclipsing RMSPs are assumed to form a common class of transitional
pulsars, this class is characterized by an spin frequency distribution
intermediate between faster nuclear MSPs and slower non-eclipsing
radio pulsars.

The frequency difference between nuclear MSPs and RMSPs ($\ga 230$ Hz)
cannot be due to the weak magneto-dipole spin-down torque that acts
onto the NS during the rotation-powered pulsar phase because this
develops on a long time-scale, $\tau_{em}\approx 10$~Gyr (for an NS
spinning at 500 Hz, with a magnetic dipole of $10^{26}$ G cm$^{3}$).
The spin-down torque that developes during the latest stages of the
accretion phase, caused by the interaction between the NS magnetic
field and the plasma rotating in the disc, is much stronger \citep[up
  to a factor of $\sim 100$;][]{ghosh1979}. As this torque is more
intense when the NS is faster ($\propto \nu^{2}$, e.g.,
\citealt{rappaport2004}), it eventually balances the rate at which the
accreted matter delivers its specific angular momentum to the NS
($\propto \nu^{-1/3}$). This allows us to define an equilibrium period
at which the overall torque acting onto the NS vanishes, the exact
value of which depends on assumptions on the disc structure and how
the field is twisted by the differential rotation of the disc plasma
(\citealt{davidson1973,ghosh1979,wang1987,bozzo2009}; see, e.g.,
\citealt{ghosh2007} for a review). Here, we considered the estimate
given by \citet{kluzniak2007} for a 1.4 $M_{\odot}$ NS accreting from
an optically thick, gas-pressure-dominated disc, as
\begin{equation} \label{eq:magequil}
  \nu_{\rm eq}^{\rm (magn)}\simeq 723 \: \mu_{26}^{-6/7}
  \:\dot{M}_{-10}^{3/7}\:\rm{Hz},
\end{equation} 
where $\mu_{26}$ is the NS magnetic dipole in units of $10^{26}$ G
cm$^3$, and $\dot{M}_{-10}$ is the mass accretion rate in units of
$10^{-10}$ M$_{\odot}$ yr$^{-1}$. We note that this estimate does not
take into account the possible ejection of matter by the rotating
magnetosphere (see below). At the end of the accretion stage, the
companion star decouples from its Roche lobe over a typical time-scale
of $\approx 100$ Myr \citep{tauris2012}.  If the NS would spin at the
magnetic equilibrium until the complete turn-off of mass accretion,
this would result in a spin-down of the NS well below the observed
spin frequencies \citep{ruderman1989}. A departure from spin
equilibrium must therefore occur at some point. \citet{tauris2012}
showed that the spin equilibrium is broken by the rapidly expanding
magnetosphere resulting from the significant decrease in the ram
pressure of the transferred material when the donor star terminates
its mass-transfer phase. The resulting spin-down is aided by the onset
of a propeller phase \citep{illarionov1975} that acts to bring the NS
farther out of equilibrium during the Roche-lobe decoupling phase.

Depending mainly on the magnetic field strength of the accreting
pulsar, the loss of rotational energy computed by \citet{tauris2012}
typically amounted to about 50~per~cent, corresponding to a decrease
in spin frequency of a factor $\sqrt{2}$, when the recycled pulsars
switch from the accretor to the final long-term rotation-powered
state. This value is of the order of the difference observed between
the spin distribution of nuclear MSPs and RMSPs, making the spin-down
experienced by the NS during the permanent decoupling of the
companion star from its Roche lobe a viable explanation.

The spin distribution of accreting MSPs, on the other hand, is closer
to that of RMSPs than to nuclear MSPs. This might originate from the
differences between the two classes of accretion-powered sources.
Unlike nuclear MSPs, accreting MSPs possess a magnetosphere that
during the quiescent states may push away the in-flowing matter,
letting the NS turn on as a rotation-powered radio pulsar \citep[{\it
    the radio-ejection
    mechanism},][]{kluzniak1988,stella1994,campana1998,burderi2001,papitto2013nat}.
Because these accreting MSPs spend most of the time in quiescence,
their spin evolution is mainly driven by the magneto-dipole torque,
and not by stronger torques associated with disc-field
interactions. This implies a slower spin evolution than nuclear MSPs,
so that a smaller spin difference might be expected between
accreting MSPs and their rotation-powered descendants. The only
accreting MSP from which many cycles between outbursts and quiescence
have been observed, SAX J1808.4-3658 \citep[see][and references
  therein]{patruno2012b}, spins down at a rate corresponding to a
slow-down by just $\sim\!5 $ Hz over the typical time-scale (100~Myr)
associated with the Roche-lobe decoupling phase of more massive donor
stars that result in  helium white dwarf companions.

\begin{figure}
\includegraphics[height=6.5cm]{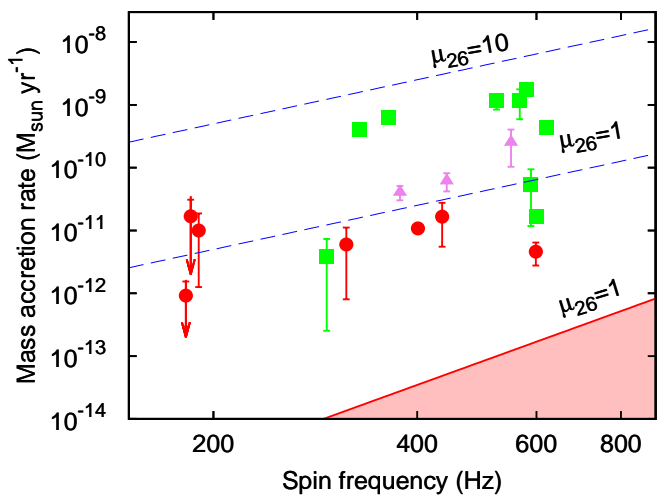}
\caption{Long-term mass-accretion rate of millisecond pulsars in the
  accretion phase, as evaluated by \citet{watts2008}. Accreting-MSPs
  are plotted as red circles, nuclear MSPs as green
  squares. Magenta triangles mark the accreting MSPs that showed
  pulsations only intermittently, and that may have a magnetic
  field close to the threshold for field burial \citep{cumming2008}.
  Blue dashed lines represent the magnetic spin equilibrium lines
  defined by Eq.\ref{eq:magequil},  evaluated for $\mu_{26}=1$ and
  $\mu_{26}=10$. The red shaded region defines the
  parameter space in which a source enters the radio ejector state for
  $\mu_{26}=1$ \citep[e.g.,][]{burderi2001}.}
 \label{fig:mdot}
\end{figure}

Our analysis did not identify a statistically significant difference
between the spin distributions of accreting and nuclear MSPs, although
the average frequency shown by the latter class is higher by
$\sim\!150$~Hz.  However, the differences between these two LMXB
classes and the much larger sample of RMSPs can be taken as an
indication that a difference  between accreting and nuclear MSPs
may exist that currently cannot be detected because of the few known
objects. A similar conclusion is reached from the significant
difference found (95~per~cent confidence level) from comparing nuclear
MSPs with the whole sample of transitional MSPs and not only accreting
MSPs. If nuclear and accreting MSPs are assumed to spin at or close to
the equilibrium value given by Eq.~\ref{eq:magequil}, the contrast
between the average frequencies of the two samples can be explained by
a difference in the mass-accretion rate of $\Delta\dot{M}/\dot{M}\sim
0.8$, and/or in the dipole moment of $\Delta\mu/\mu\sim
-0.4$. Nuclear MSPs generally accrete at a higher average rate than
accreting MSPs (see green squares and red circles in
Fig.~\ref{fig:mdot}), and most probably possess a weaker magnetic
field because they do not show coherent pulsations. A weaker field
might also reflect a higher mass-accretion rate as well because a
plasma in-flowing at a high rate diamagnetically screens the magnetic
field of the NS below its surface
\citep{bhattacharya1995,cumming2001,cumming2008}.  These
considerations may be taken as a clue that nuclear MSPs are found in
an earlier evolutionary stage that is characterized by a higher
average accretion rate (and a consequently lower dipolar magnetic
field) and a faster rotation than to accreting MSPs. The latter are
instead observed in a phase much closer to the accretion turn-off and
to the switch-on of an RMSP, as also demonstrated by the rapid changes
of state observed in some of them.

An important role in explaining the different observed spin
distributions of some classes might be played by their different
evolutionary history. All accreting- and nuclear MSPs for which the
orbital period is known belong to systems with $P_{orb}<1$~day.  These
binaries originate in systems in which the companion star had a low
mass (1-$2\;M_{\odot}$) and was still on the main sequence, at the
onset of mass transfer.  The same holds for eclipsing RMSPs, that is,
the so-called black widows and redbacks, \citep[cf.][and references
  therein]{chen2013}.  In contrast, RMSPs originate in a wider sample
of binaries, comprising both systems with low-mass \citep{ts99} and
intermediate-mass ($2-6\;M_{\odot}$) companions \citep{tvs00}.
Furthermore, depending on their final $P_{orb}$ and the composition of
the white dwarf companion, in some of these systems the donor star had
already evolved off the main sequence at the onset of the mass
transfer.  Hence, all RMSPs with massive white dwarf companions and
RMSPs with helium white dwarf companions in wide orbits ($P_{orb}\gg
1\;{\rm day}$) are only expected to be mildly recycled as a
consequence of the relatively short-lived mass-transfer phase in these
systems \citep[see Figs.~1 and 9 in][]{tauris2012b}.  Given that the
observed sample of non-eclipsing RMSPs contains a fraction of such
mildly recycled (relatively slow spinning) sources might therefore
partly explain why their observed spin distribution is slightly slower
than that of eclipsing RMSPs.  However, this selection bias probably
does not affect the comparison between nuclear and transitional MSPs
because all these known sources evolved from low-mass companions in
short ($P_{orb}<1$~day) orbital period systems, ensuring a long phase
of mass transfer and efficient recycling.  Note that an expected short
and intense mass-transfer phase in wide-orbit systems can explain the
hitherto non-detection of accreting- and nuclear MSPs in such binaries
as well.

 On the other hand, quantifying the impact of instrumental selection
 biases is more complex and would deserve a population synthesis
 analysis, which is beyond the scope of this paper. Nevertheless, some
 conclusions can be drawn on the basis of the observed properties of
 MSPs in light of the searching strategies actually employed to detect
 them.  Frequency Doppler shifts induced by the orbital motion limit
 the time interval $T_{\rm best}$ over which a search for a coherent
 signal can be performed \citep[$T_{\rm best}\propto P_{orb}^{\,2/3}$;
   see Eq. 21 in][]{johnston1991}. This significantly reduces the
 minimum detectable amplitude $A_{min}$ of accretion powered X-ray
 pulsations emitted by NS in very-short period binaries ($P_{orb}\la$
 1~hr)\footnote{For a 600 Hz pulsar in a $P_{orb}=0.5$~hr binary
   observed at an X-ray flux of 200 counts s$^{-1}$ (typical, e.g., of
   observations performed with EPICpn on-board {\textsl XMM-Newton}),
   $T_{\rm best}$ is reduced below 30 s, a time interval that only
   allows the detection of strong signals with an amplitude larger
   than $A_{min}\simeq12$ per cent \citep{vaughan1994}, higher than
   the typical few per cent amplitude signals observed from
   accreting MSPs \citep{patruno2012}. A similar result is obtained
   ($A_{min}\simeq10$ per cent) if a 300 Hz pulsar is considered.},
 but since the dependence of this effect on the pulsar spin frequency
 is only weak ($A_{min}\propto \nu^{1/4}$), the resulting bias on the
 observed spin frequency distribution of accreting MSPs is small.  On
 the other hand, the orbital motion does not affect the detection of
 nuclear MSPs at all because the burst oscillations are observed at
 X-ray fluxes $\ga 100$ times higher than accreting MSPs signals, and
 a detection is achieved in time-intervals of a few seconds
 \citep{watts2012}, which is much shorter than the orbital period of
 the binary.  Pulse arrival-time delays induced by orbital motion and
 by propagation in the ionized interstellar medium also reduce the
 sensitivity to high-frequency signals in the radio band
 \citep{dewey1985,hessels2007}. However, recent improvements in data
 acquisition systems, computational power, and search algorithms have
 reduced these effects, resulting in doubling the detections of RMSPs
 in the past five years \citep{lorimer2013,ransom2013}. This increase
 is also caused by radio follow-up of unidentified $\gamma$-ray {\it
   Fermi}-LAT sources, which now amount to $15$ per cent of the total
 number of RMSPs \citep{ray2012,ransom2013}. As the $\gamma$-ray
 output of a millisecond pulsar increases with its spin-down power
 ($L_{\gamma}\propto \dot{E}_{sd}^{a}$, with $a=0.5$--$1$ and
 $\dot{E}_{sd}\propto \mu^2\nu^4$; see \citealt{abdo2013}), the
 detection of fast-spinning sources through this method is
 favoured. During the past five years, the follow-up of {\it Fermi}
 sources determined an eight-fold increase in the number of known
 Galactic eclipsing RMSP \citep{roberts2013}, which are significantly
 faster than the rest of RMSP, as we have shown in this work.  Before
 this overall reduction of the bias against the detection of fast
 radio pulsars, \citet{lorimer2013} estimated the peak of the spin
 frequency distribution of RMSP of the Galactic plane to be
 $\Delta\nu\simeq65$ Hz higher than the observed average. This rapid
 increase of the MSPs population over the past years then indicates
 that even if a complete sample of fast pulsars has not been achieved
 yet, the impact of selection bias has been reduced below this
 level, $\Delta\nu$.

In the future, X-ray observatories such as the {\it Large Observatory for
  X-ray Timing} \citep[LOFT][]{feroci2012}, and to some extent {\it
  AstroSAT} \citep{agrawal2006} and {\it NICER} \citep{gendreau2012},
are expected to increase the sensitivity to fast coherent signals from
accreting sources up to a factor $\sim 5$. At the same time, further
improvement in the computational power and analysis techniques used to
discover RMSPs, the numerous candidates identified by {\it Fermi}-LAT,
and the advent of the {\it Square Kilometre Array}
\citep[SKA][]{smits2009} will also increase the number of fast radio
pulsars in close binary systems. An increase in the number of sources
known will be fundamental to constrain the details of pulsar recycling
from the point of view of the observed spin distributions.

\begin{acknowledgements}

We acknowledge grants AYA2012-39303, SGR2009- 811, and
iLINK2011-0303. AP is supported by a Juan de la Cierva fellowship. NR
is supported by a Ramon y Cajal fellowship and by an NWO Vidi
Award. DFT is additionally supported by a Friedrich Wilhelm Bessel
Award of the Alexander von Humboldt Foundation.  T.M.T. gratefully
acknowledges financial support and hospitality at both the
Argelander-Insitut f\"ur Astronomie, Universit\"at Bonn and the
Max-Planck-Institut f\"ur Radioastronomie.

\end{acknowledgements}

\bibliographystyle{aa}
\bibliography{biblio}

\end{document}